# Absolute quantum gravimeters and gradiometers for field measurements

L. Antoni-Micollier[1], M. Arnal[1], R. Gautier[1], C. Janvier[1,2], V. Ménoret[1,2], J. Richard[1], P. Vermeulen[1], P. Rosenbusch[1], C. Majek[1], B. Desruelle[3]

[1] Exail Quantum Systems, Institut d'Optique d'Aquitaine, rue François Mitterrand 33400 Talence, France

[2] Laboratoire Photonique, Numérique et Nanosciences (LP2N), Université Bordeaux – IOGS – CNRS, 1 rue François Mitterrand, 33400 Talence, France

[3] Exail, 34 rue de la Croix de Fer, 78100 Saint Germain-en-Laye, France

Gravity measurements provide valuable information on the mass distribution below the earth surface relevant to various areas of geosciences such as hydrology, geodesy, geophysics, volcanology, and natural resources management. During the past decades, the needs for sensitivity, robustness, compactness, and transportability of instruments measuring the gravitational acceleration have constantly increased. Today, applications typically call for 1 µGal = 10 nm/s$^2$ ~10$^{-9}$ g resolution on time scales ranging from minutes to years. Absolute Quantum Gravimeters (AQGs) based on matter-wave interferometry with laser-cooled atoms address all these challenges at once, even in uncontrolled environments [1, 2]. Furthermore, to date, quantum gravimeters are the only technology capable of providing continuous absolute gravity data over long measurement durations (~1 day to months or more). In this paper we recall the AQG working principle and present the reproducible high performance at the µGal level on all 16 units fabricated so far. We also describe recent progress on the Differential Quantum Gravimeter (DQG) which measures simultaneously the mean gravitational acceleration and its vertical gradient at the level of 10 nm/s$^2$ and 1 E (1 E[eotvos] = 10$^{-9}$ s$^{-2}$) respectively [3].

The first absolute gravimeter based on quantum technology was realised in 1992 [4]. Rather than a macroscopic object, laser-cooled atoms in ultra-high vacuum are the falling test mass. This allows both for a near-perfect free-fall trajectory and for continuous long-term operation because the system has no moving parts. Pulses of vertically propagating laser light manipulate the quantum state of the atoms and thereby open a matterwave interferometer that is sensitive to the vertical acceleration caused by the Earth gravity attraction.

Since this first laboratory demonstration, scientific and technological progress have led to considerable improvements in maturity for atom-interferometers in general [5] and absolute gravimeters in particular [1, 6]. Absolute gravimeters based on quantum technology have gained in compactness and robustness, while keeping attractive specifications [7, 8, 9]. These prototypes were realized as part of research projects and none has been fabricated under an industrial approach, where reliability and reproducibility are the prime criteria.

In this paper, we present the industrial fabrication of Absolute Quantum Gravimeters based on laser-cooled atoms [1]. We show the reproducible high performance of all 16 units fabricated in our facilities so far, both for short-term sensitivity and long-term stability at the 10 nm/s$^2$ level, verified by our thorough evaluation process prior to delivery. We furthermore present recent results obtained with our industrial prototype Differential Quantum Gravimeter (DQG). Its working principle can be seen as two absolute quantum gravimeters stacked vertically within the same set-up, sharing the same vertical laser beam. Common-mode measurement yields the value of gravity, and in differential mode the instrument measures the vertical gradient of gravity with the benefit of rejecting common-mode noise sources [3]. Despite its simple architecture, the device can measure both quantities with excellent resolution: gravity measurements are comparable with those of the AQG and the differential measurement is limited by quantum projection noise, which is the fundamental physics limit for such quantum sensors. Thereby, the DQG reaches a long-term stability of $0.15$ E $= 1.5 \ 10^{-10}$ s$^{-2}$. This novel type of sensor offers promising applications in the detection of shallow anomalies such as cavities or high-density structures, as well as in hydrology and volcanology.

**The Absolute Quantum Gravimeter (AQG)**
The AQG is a commercial product that complies with the characteristic requirements of field operation: high transportability, compacity, robustness and adaptation to the vibration spectrum of the measurement site. Our more than 10 years technology development have brought optics, electronics and vacuum components to the autonomy and robustness necessary for field deployment. Thereby, the AQG features low maintenance and ease of use.

The AQG operates in a cyclic fashion, where one cycle comprises four main steps, during which the atoms are trapped, prepared in a specific quantum state, interrogated via a matterwave interferometer, and detected (Fig. 1). All steps are realized through one unique laser beam [1]. From the detection signal, a raw gravity value is calculated at each cycle, with a repetition rate of nearly 2 Hz. The raw value is then corrected from environmental effects, such as Earth tides, polar motion and atmospheric pressure, together with instrumental parameters such as tilt, laser wavelength and reference-quartz frequency.

The main noise contribution in the AQG signal is the effect of ground vibrations whose vertical component can fundamentally not be distinguished from the static gravitational acceleration. We mitigated this noise by an active compensation system [7] where ground vibrations are measured by a classical accelerometer and fed-back to the laser system, so that the atoms experience an undisturbed optical phase, even when the instrument is subject to vibrations. We can adapt this compensation system to the characteristic vibration spectrum of the measurement site both by selecting the optimal classical sensor and adapting the compensation parameters [2]. We have furthermore checked that the compensation system does not affect the mean absolute value of g. When installed outside, wind can create vibrations, too, that appear as additional noise if not entirely rejected by the compensation system. Installation of a leisure tent has been an effective wind protection.

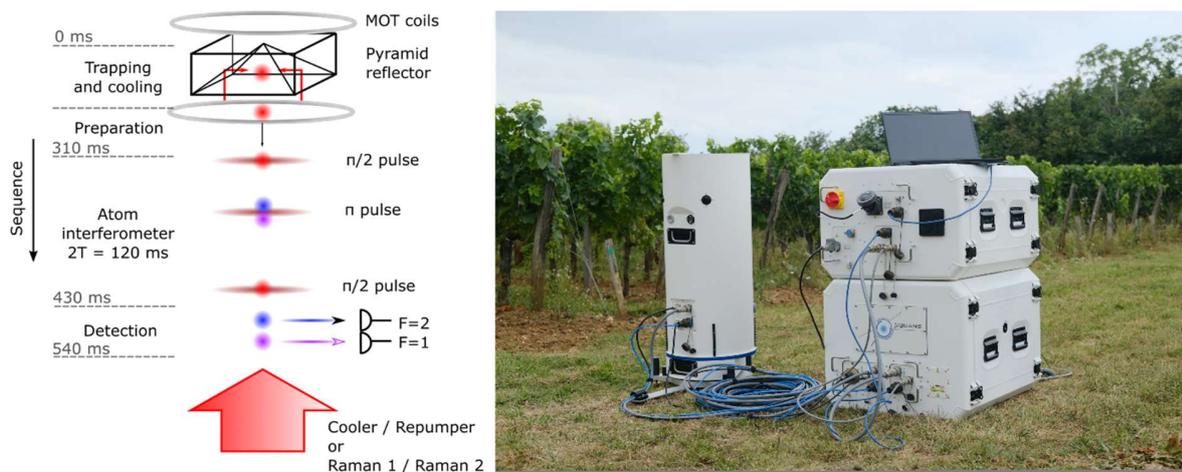

***Fig. 1.*** Left: sketch of the sensor head and quantum gravimeter measurement principle, adapted from [1]: a cloud of ~$10^6$ Rb atoms is prepared in a magneto-optical trap (MOT) formed by a laser beam and its reflections in a hollow pyramid reflector. After 270 ms of trapping, the light is switched off so that the atoms begin to fall. A short preparation stage selects the atoms with the required internal quantum state, before, during the atom free fall, 3 laser pulses drive a matterwave interferometer that is sensitive to the vertical acceleration. Detection of the number of atoms in each interferometer output port, labelled by their internal quantum state, allows to calculate the experienced acceleration. Right: Picture of the field version of the Absolute Quantum Gravimeter. It comprises 3 modules, the sensor head (cylinder on the left), the power supply (top module on the right) and the laser and electronics rack (bottom module), each weighing less than 40 kg in a volume of 300 L or less each.

The AQG is fabricated in two versions: the AQG-A version is for laboratory use in stable temperature conditions. The AQG-B with integrated temperature stabilization is designed for outdoor use. It is composed of three main modules (Fig. 1): (i) a power supply module delivering DC voltages and the thermal regulation, (ii) the laser system and control electronics, and (iii) the sensor head where the gravity measurement takes place. Thanks to the thermal control, the device operates in temperatures ranging from 0 to 40°C. Specific versions for cold climate (-10°C) have been realized on customer request. To ease transportation, installation and packing, each module has a compact size of ≤ 300 L and < 40 kg weight. A 15 m long cable bundle connects the electronics to the sensor head, so that the head can be moved within a certain perimeter without displacing the electronics. The number of connections to the sensor head has been reduced to 4, including one single optical fibre. The electric power consumption is 500 W for the entire system. This may be lowered in future AQG versions by reducing the demand of temperature stabilisation.

*Repeatedly reliable performance*
Since 2018, Exail has produced, tested, and delivered to users worldwide 4 AQG-A and 12 field units (AQG-B). The continuous operation of the first AQG-B [10] since 2019 has validated the field-compatible robustness of the technical subsystems and components, control software and gravity correction algorithm. This and all following units were functionally tested at our factory in suburban Bordeaux, France. Their metrological performances were also validated prior to shipment. No difference of performance has been observed between the laboratory and field version, despite their hardware differences. In particular, the short-term gravity sensitivity of all units is systematically better than 800 nm/s$^2$/$\tau^{1/2}$ as can be seen from the Allan deviations plotted in Fig. 2. The measurements are limited by the high level of microseismic noise at our workshop on the second floor of an inner-city building and its proximity to a tramway and the ocean. The limitations of our test site become obvious when comparing the evaluation of AQG B10, which gave a sensitivity of 770 nm/s$^2$/τ$^{1/2}$ at the factory (orange line in Fig 2) to a measurement of the same instrument at the quiet geodetic observatory in Wettzell, Germany, where the sensitivity was 430 nm/s$^2$/τ$^{1/2}$ (red line in Fig 2). Furthermore, for AQG A02, which showed typical sensitivity at our factory, a selected data set at the gravimetric reference station Bad Homburg, Germany shows a sensitivity of 300 nm/s$^2$/τ$^{1/2}$ during the night of 30/10/2020 to 01/11/2020, the 1$^{st}$ November being a bank holiday in Germany (blue line in Fig 2). This sensitivity corresponds to the technical AQG laser phase noise, which is determined by the hardware choices. The long-term stability below 10 nm/s$^2$ is confirmed for all AQG independent of the site. These results confirm our ability to fabricate quantum gravimeters with reproducible excellent performances.

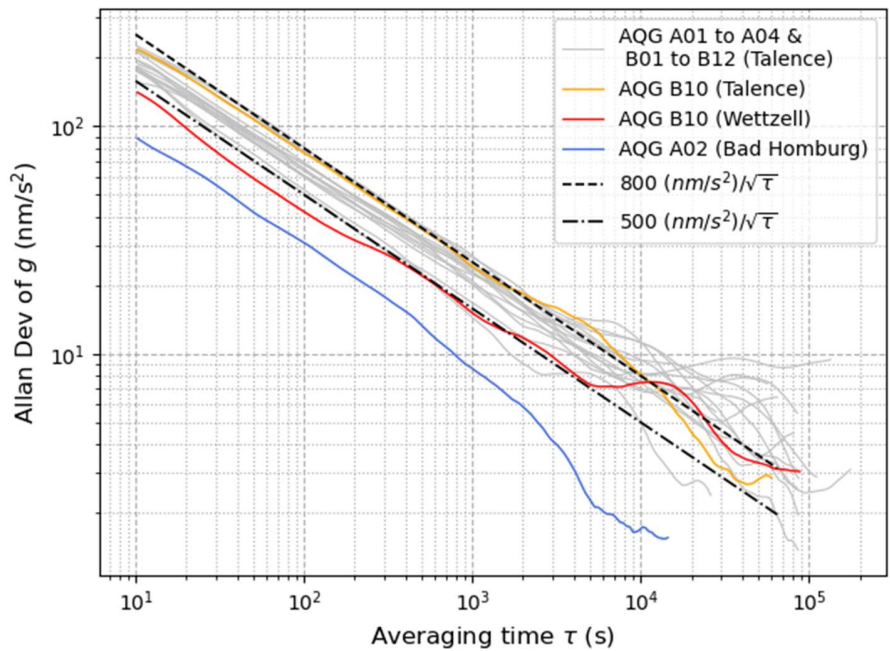

*Fig. 2.* Allan deviation of the corrected gravity data recorded at our factory in Talence, suburban Bordeaux, by all Absolute Quantum Gravimeters produced so far, AQG A01 to A04 and B01 to B12 (grey lines, data of B10 highlighted in yellow). For all units the short-term sensitivity is below 800 nm/s$^2$/τ$^{1/2}$. It is limited by the floor vibrations in our laboratory on the 2nd level of a 4-story building. Measurements with AQG B10 at the geodetic observatory Wettzell, Germany confirm the instrument sensitivity below 500 nm/s$^2$/τ$^{1/2}$ (red line, data courtesy of J. Glässel and H. Wziontek, BKG). Selected data of AQG A02 at the gravimetric reference station Bad Homburg, Germany demonstrate an instrument sensitivity of 300 nm/s$^2$/τ$^{1/2}$ (blue line, data courtesy of J. Glässel and R. Falk, BKG). The long-term stability is below 10 nm/s$^2$ for all AQG independent from the site.

### *A 3-year long field campaign*

Among the 16 quantum gravimeters delivered to date, we here focus on AQG-B03, which is installed close to the summit craters of Mt. Etna in Sicily since August 2020. In the framework of the European "NEWTON-g" project, the INGV operates the instrument in this harsh environment, where strong daily and seasonal temperature variations, volcanic tremor, lava fountain episodes, and corrosive volcanic gases are present together with frequent power-cuts [2,11].

Figure 3 shows gravity data recorded by AQG-B03 between August 2020 and October 2021. Small gaps in the time series are due to temporary failures of the off-grid power supply system constituted of solar cells and a diesel generator. During winter 2021, when the generator failed and snow made the site inaccessible, the AQG fell into unwanted hibernation without electricity by hence lacking both temperature stabilisation and active vacuum pumping. This occurred repeatedly over the following two winter seasons. After each of these long power failures, a remote restart of all subsystems (ion pump, electronics, lasers) was possible and data acquisition could resume within a few hours from power-on.

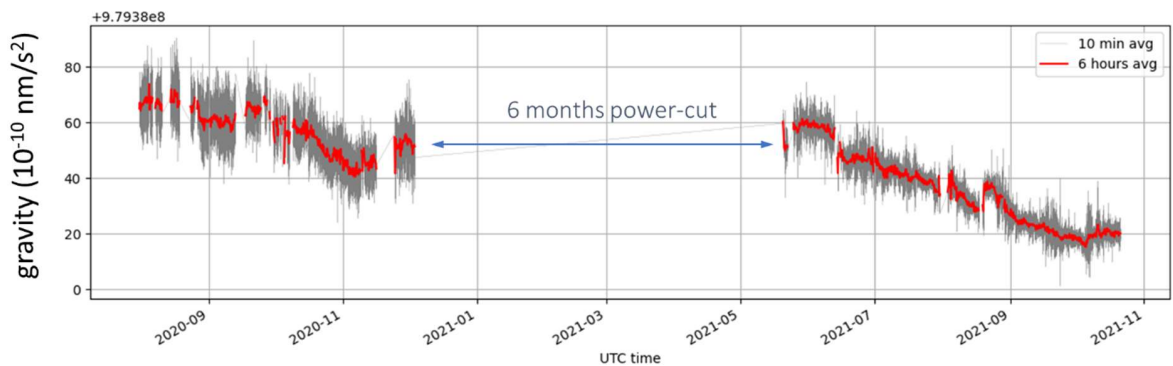

*Fig. 3.* AQG-B03 gravity time series from August 2020 to October 2021 produced in the framework of the NEWTON-g project on Mt Etna, Sicily. Environmental effects including Earth tides have been removed. The data are averaged over 10 min (gray curve) and 6 h (red curve) intervals. Despite the strong volcanic tremor and frequent power-cuts, including the 6 months winter span, the AQG reliably produced relevant data.

**The Differential Quantum Gravimeter (DQG)**
The DQG is an instrument capable of performing measurements of both gravity and its vertical gradient within a compact and transportable design. Based on the same technological bricks as the AQG, it uses two vertically stacked identical atom interferometers that measure gravity at two different heights while sharing the same interrogation laser, vacuum chamber and reference mirror (Fig 5, left). Both top and bottom clouds share the same measurement sequence, which is similar to that of the AQG, the only significant difference being a longer interferometer duration of 240 ms instead of 120 ms for an AQG and a consequently longer cycle duration of 1.08 s [3]. Two pyramidal reflectors are positioned on top of each other, so that each step of the sequence is performed simultaneously for the two interferometers. In particular, the simultaneous interrogation and detection of the two atom clouds by the same laser ensures the best common-mode rejection of ground vibrations.

Another difference from the AQG is the addition of a piezoelectric tip-tilt platform under the DQG reference mirror. This platform rotates the mirror during the measurement on a trajectory opposite to the Earth rotation and enables to cancel any Coriolis effect coming from residual eastward/westward velocities of the two atom clouds. The platform was implemented to comply to the more stringent requirements on the atom velocities (about 10 µm/s differential velocity) to ensure the stability of the gradient measurement.

The phase of each matterwave interferometer represents the gravitational acceleration integrated over the respective atom trajectory. As for the AQG, the sensitivity to variations of this phase is maximized by interrogating of the interferometers at midfringe. In the case of the DQG, a dual feedback loop keeps both interferometers at midfringe, and maximizes sensitivity to gravity and vertical gradient. This feedback is performed by retroacting on the frequency of the Raman lasers, which allows for gradient measurements without requiring a precise and continuous measurement of the distance between the two atomic clouds [12].

The DQG in its current laboratory version is composed of two subsystems (Fig. 5) with reduced size, weight, and power consumption. The sensor head (175 cm high, 75 kg) contains the vacuum chamber where the measurement takes place. The electronics and laser module (0.1 m$^3$, 33 kg) generates all optical and electrical signals necessary for the instrument control and atom manipulation. The overall power consumption is approximately 200 W.

The DQG is a versatile instrument which can be used both for long, temporal measurements and for short spatial surveys. These two types of measurements are discussed in the following paragraphs.

*Observatory type measurements*
As an example of temporal measurements, we present a 21-day acquisition that has been recorded by the DQG under laboratory conditions. (Fig. 4). The raw gravity signal follows remarkably well the tide model of our factory site. Calculating the residuals shows a peak-to-peak variation of less than 50 nm/s². Between 2021/11/11 and 2021/11/16, one observes very little noise, because this period corresponds to a series of non-working days resulting in a lower ground vibration noise and a cleaner gravity signal. Note that this calm period is not visible in the gravity gradient signal because vibrations affect the two clouds in common-mode and are always eliminated by the differential measurement. The Allan deviation of the gravity signal demonstrates a sensitivity of 650 nm/s²/$\tau^{1/2}$ and a long-term stability around 5 nm/s². The Allan deviation of the gravity gradient demonstrates a sensitivity of 56 E/$\tau^{1/2}$ and a long-term stability around 1 E.

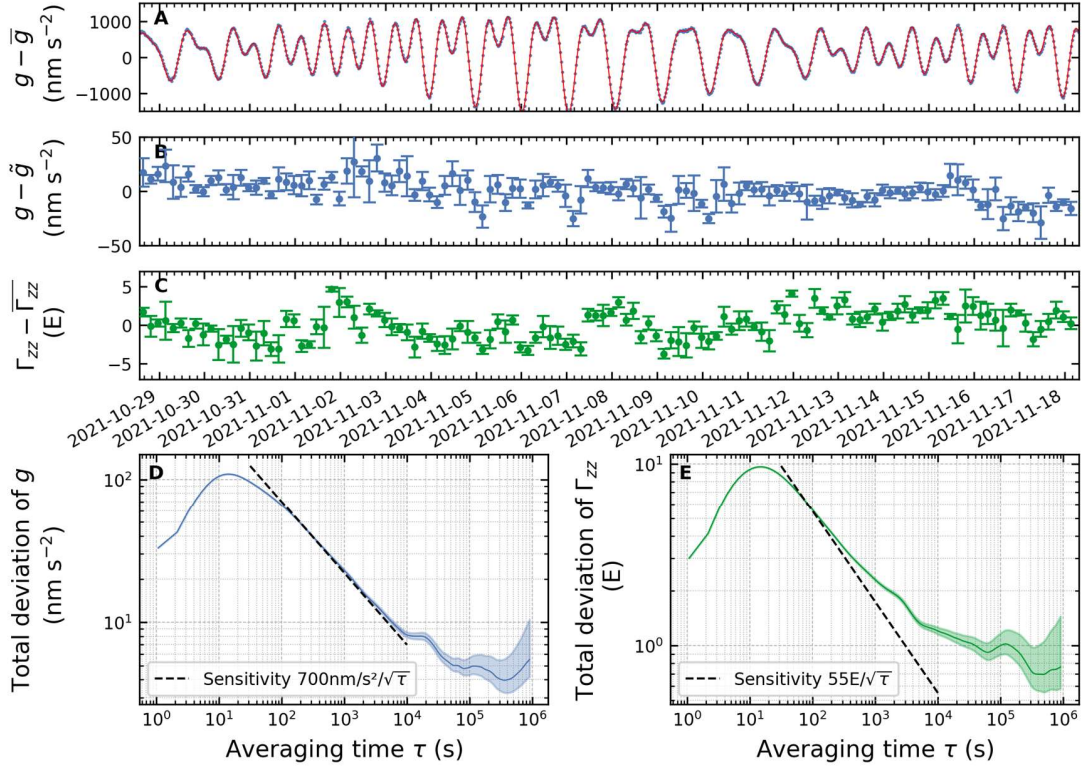

*Fig. 4.* Gravity data recorded by the DQG at our factory site over 21 days. A) Gravity data corrected from instrument tilt and atmospheric pressure (blue dots), together with the tides model (red line). B) gravity residuals after subtraction of the tides model. C) Gravity gradient data, without any correction. D) Allan deviation of the gravity residuals. E) Allan deviation of the gravity gradient signal. The sensitivity of gravity measurement is 650 nm/s²/$\tau^{1/2}$ and 56 E/$\tau^{1/2}$ for the gravity gradient.

*Survey measurements*

As an example of spatially resolved data, we performed a short survey in the car park of our factory. Two 1 m³ cavities are buried just beneath the asphalt and covered by two cast iron manholes. We took a series of 30 minutes measurements at different positions on a line passing over one of the cavities. The data are shown in figure 5. We observe a clear signal on the gravity gradient which is well reproduced by a crude model of the two cavities (black line). On the other hand, the gravity data seem to be dominated by noise, although there may be an oscillating signal that is not accounted for by our simple model. It may arise from the combination of the cavities with the close-by building and cm-scale terrain elevation variations. A full terrain model would be needed here.

The survey shows that the DQG is able to detect small nearby gravity anomalies with a better signal-to-noise-ratio than gravity. Furthermore, the simultaneous measurement of both quantities opens new horizons for gravity inversion using the different signatures observed for gravity and the gravity gradient. The signature difference will indicate if the anomaly is small, or large, due to deep or shallow sources.

In its current form the DQG is a laboratory prototype with wheels for manipulation on flat, hard ground. The on-going Horizon Europe project FIQUgS is dedicated to the development of a more compact, field compatible DQG which will be moved by a robot on natural terrain. The compactness comes at a cost of sensitivity, but mitigation strategies are under development, too.

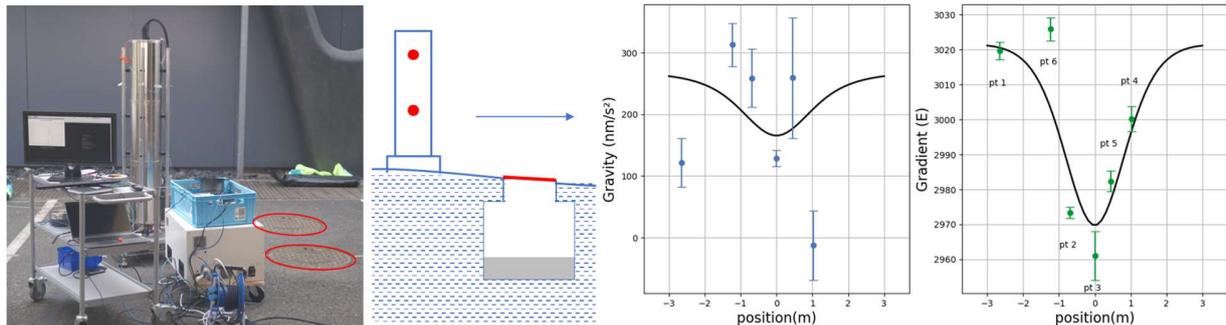

*Fig. 5.* From left to right: Picture of the DQG during a ground survey which moved it across two underground cavities (red circles). The sensor head (metallic cylinder) is moved in steps of about 0.5 m along a line crossing the cavities. The laser system (white box) was placed on a skateboard to follow the head. A schematic of the measurement set-up shows the sensor head with two vertically stacked atom clouds and one of the cavities of roughly 1 m³ volume. The gravity measurement (blue points) does not allow to detect the cavity, but the gravity gradient shows a clear signature, which coincides with a simple model based on the cavity volume and position (black line).

**Conclusion**

Absolute gravimeters based on quantum technology use laser-cooled atoms as falling test mass and are thereby free of moving parts. This exalts the instrument robustness, high measurement rate and low maintenance. Since 2018, Exail commercializes absolute quantum gravimeters, and all 16 fabricated units show a vibration limited sensitivity better than 800 nm/s²/$\tau^{1/2}$ at our inner-city factory site. At a dedicated geodetic site, it is typically better than 500 nm/s²/$\tau^{1/2}$. The long-term sensitivity is better than 10 nm/s² = 1 µGal. The evaluation of the AQG accuracy will be the subject of a forthcoming publication. A strong demonstration of the robustness of the AQG is the unit which has been operating on Mt Etna for over three years now. Exail's differential quantum gravimeter (DQG) which measures the mean gravity value and its vertical gravity gradient simultaneously, reaches similar performances for the mean gravity value and a sensitivity of about 60 E/ $\tau^{1/2}$ for the gradient with a long-term stability around 1 E. On-going R&D aims towards commercialization as well as the development of a field version that can be mounted on an autonomous rover.

**Acknowledgment**
The authors thank the technical, administrative and scientific staff of Muquans/iXblue/Exail for their work and support. We acknowledge the INGV-OE technical and scientific staff for their support in maintaining the facilities at the PDN observatory. We thank J. Glässel, R. Falk and H. Wziontek for sharing the Wettzell and Bad Homburg data with us. This work was supported by the NEWTON-g project, which has received funding from the EC Horizon 2020 program under the FETOPEN-2016/2017 call (Grant Agreement No 801221) and the EC Horizon-Europe program FIQUgS (Grant Agreement No 101080144). We acknowledge financial support by the ANR under Contract No. ANR-19-CE47-0003 GRADUS. C. Janvier and V. Ménoret acknowledge financial support by ANR France Relance.

**References**
[1] V. Ménoret, P. Vermeulen, N. Le Moigne, S. Bonvalot, *et al.* Gravity measurements below $10^{-9}$ *g* with a transportable absolute quantum gravimeter. *Sci Rep* **8**, 12300 (2018).

[2] L. Antoni-Micollier, D. Carbone, V. Ménoret, J. Lautier-Gaud, T. King, *et al.* Detecting volcano-related underground mass changes with a quantum gravimeter. *Geophysical Research Letters*, 2022, vol. 49, no 13, p. e2022GL097814.

[3] C. Janvier, V. Ménoret, B. Desruelle, S. Merlet, A. Landragin, F. Pereira dos Santos. Compact differential gravimeter at the quantum projection-noise limit. *Physical Review A*, 2022, vol. 105, no 2, p. 022801.

[4] M. Kasevich, S. Chu. Measurement of the gravitational acceleration of an atom with a light-pulse atom interferometer. *Applied Physics B*, 1992, vol. 54, p. 321-332.

[5] R. Geiger, A. Landragin, S. Merlet, F. Pereira Dos Santos. High-accuracy inertial measurements with cold-atom sensors. *AVS Quantum Science*, 2020, vol. 2, no 2.

[6] F. Pereira Dos Santos, and S. Bonvalot. Cold-Atom Absolute Gravimetry, In *Encyclopedia of Geodesy*, (ed. Grafarend, E.) 1–6 (Springer International Publishing, 2016).

[7] J. Lautier, L. Volodimer, T. Hardin, S. Merlet, M. Lours, *et al.* Hybridizing matter-wave and classical accelerometers. *Applied Physics Letters*, 2014, vol. 105, no 14.

[8] C. Freier, M. Hauth, V. Schkolnik, B. Leykauf, M. Schilling, *et al.* Mobile quantum gravity sensor with unprecedented stability. *Journal of physics: conference series*. IOP Publishing, 2016. p. 012050.

[9] Y. Bidel, N. Zahzam, C. Blanchard, A. Bonnin, M. Cadoret, *et al.* Absolute marine gravimetry with matter-wave interferometry. *Nature communications*, 2018, vol. 9, no 1, p. 627.

[10] A.-K. Cooke, C. Champollion, and N. Le Moigne. Evaluation of the capacities of a field absolute quantum gravimeter (AQG# B01). *Geoscientific Instrumentation, Methods and Data Systems Discussions*, 2020, vol. 2020, p. 1-24.

[11] D. Carbone, L. Antoni-Micollier, G. Hammond, E. de Zeeuw-van Dalfsen, E. Rivalta, *et al.* The NEWTON-g gravity imager: Toward new paradigms for terrain gravimetry. *Frontiers in Earth Science*, 2020, vol. 8, p. 573396.

[12] R. Caldani, K.X. Weng, S. Merlet, F. Pereira Dos Santos. Simultaneous accurate determination of both gravity and its vertical gradient. *Physical Review A*, 2019, vol. 99, no 3, p. 033601.